\documentclass[12pt,amssymb,epsfig]{article}

\usepackage{amssymb}
\usepackage{epsfig}

\include{figures}
\newcommand{\simgt}
{\mbox{\raisebox{-0.5ex}{$\textstyle \; \sim$}
\raisebox{ 0.8ex}{$\textstyle  \!\!\!\!\!\!\! >$  }}}
\newcommand{\simr}
{\mbox{\raisebox{-0.05ex}{$\textstyle r$}
\raisebox{ 1.35ex}{$\textstyle \!\!\!\!\!\! \sim$  }}}
\newcommand{\simv}
{\mbox{\raisebox{-0.05ex}{$\textstyle v$}
\raisebox{ 1.25ex}{$\textstyle \!\!\!\!\!\! \sim$  }}}
\newcommand{\simM}
{\mbox{\raisebox{-0.05ex}{$\textstyle \; M$}
\raisebox{ 1.75ex}{$\textstyle \!\!\!\!\!\! \sim$  }}}
\begin{document}

\vspace*{2.75cm}

\footnotesize

\begin{flushleft}
{
{\Large \bf A symbiotic scenario for the rapid formation}\\[0.2cm]
{\Large \bf of supermassive black holes}\\[1cm] 

\mbox{} \hspace{2.15cm} {\large \bf M.C. Richter, G.B. Tupper and R.D. 
Viollier\footnote{Author to whom any correspondence should be addressed} }\\[0.3cm]

\mbox{} \hspace{2.15cm} \parbox{13cm}{\small Centre for Theoretical Physics and Astrophysics,
Department of Physics,\\ University of Cape Town, Private Bag, Rondebosch 7701, South Africa\\
{\bf E-mail:} viollier@physci.uct.ac.za} }
\end{flushleft} 
\vspace{0.5cm}
\mbox{} \hspace{2.15cm} \parbox{13cm}{{\small \bf Abstract.} The most massive black holes, lurking at the centers of large galaxies, must have formed 
less than a billion years after the big bang, as they are visible today in the form of bright quasars at  redshift larger than six. Their early appearance is mysterious, because the radiation pressure, generated by infalling ionized matter, inhibits the rapid growth of these black holes from stellar-mass black holes. It is shown that the supermassive black holes may form timeously through the accretion of predominantly degenerate sterile neutrino dark matter onto stellar-mass black holes. Our symbiotic scenario relies on the formation of, first, supermassive degenerate sterile neutrino balls through gravitational cooling and, then, stellar-mass black holes through supernova explosions of massive stars at the center of the neutrino balls. The observed lower and upper limits of the supermassive black holes are explained by the corresponding mass limits of the preformed neutrino balls. }\\[1cm]

\mbox{} \hspace{2.15cm} {\small \bf Keywords:} dark matter, neutrino properties, massive black holes\\[0.5cm]
\mbox{} \hspace{2.15cm} {\small \bf ArXiv ePrint:} ???\\
\newpage

\subsection*{1. Introduction}
Supermassive black holes 
of $\sim 3 \times 10^{9}M_{\odot}$ \cite{sch1}
are already present at redshift
$z = 6.42$ or 850 Myr after the big bang, as is evident from the recently discovered \cite{wil2} quasar
SDSS J114816.64+525150.3. 
Unlike the galaxies, which are thought to assemble through hierarchical mergers of smaller galaxies \cite{ell3}, the supermassive black holes form anti-hierarchically, i.e. the larger black holes prior to the smaller ones. 
This is supported by the observation \cite{com4} 
that the number of brighter quasars peaks at $z \sim 2$, while
that of lower-luminosity active galactic nuclei peaks at 
$z \sim 0.6$.
There is only a narrow window of about 485 Myr between the appearance of the first stars at $z \sim 11$, or 365 Myr after the big bang when reionization of the universe begins \cite{spe5}, 
and the appearance of the first quasars \cite{wil2} at $z = 6.42$ when reionization ends, during which at least some of the most massive black holes must have formed.\\[.2cm]
It has been known for more than three decades \cite{hil6} that the growth of a black hole from the $10M_{\odot}$ to the 
$3\!\times\!10^{8}M_{\odot}$ scale, through the accretion of baryonic matter radiating at the Eddington limit, with an efficiency between $\varepsilon \sim 0.1$ and $\varepsilon \sim 0.2$, would take at least 1.6 Gyr. This seems to preclude the existence of a ten times larger black hole at 
$z = 6.42$ that must have been produced in only $\sim$ 0.5 Gyr. 
Such a short accretion time can,
in the absence of frequent black hole mergers, 
only be achieved by drastically increasing either the ambient baryonic matter density, or the mass of the seed black hole, or both \cite{com4}. The baryonic 
matter density would have to be 
much larger than the average matter density within 1 pc of the
Galactic center of $\sim 10^{7} M_{\odot}/{\rm pc}^{3}$,
which is already uncomfortably large.
Alternatively, increasing the seed black hole mass by several orders of magnitude
is not attractive either, because, in spite of the ongoing intensive search, the 
evidence for intermediate mass black holes, with masses between 20 $M_{\odot}$ and 
$10^{6}M_{\odot}$, is rather weak and controversial \cite{mil7}.
Regardless of whether the supermassive black holes were produced through the accretion of baryonic matter onto stellar-mass black holes, or through binary mergers starting from stellar-mass black holes, these would inevitably
have left a trail of observable intermediate mass black holes, which ran out of baryonic matter
or black hole supplies sometime and somewhere in this universe.\\[.2cm] 
Thus apart from the easily detectable upper limit of
$M_{\rm max} \sim 3 \times 10^{9}M_{\odot}$,
there also seems to be a lower mass limit of
$M_{\rm min} \sim 10^{6}M_{\odot}$ of
the supermassive black holes. In this context, it is important to note that the
mass of the black hole at the center of M87 \cite{mac8}
is equal to that of the earliest quasar \cite{wil2} 
observed at 
$z$ = 6.42. As these are both archetypical examples of the most massive black holes at vastly different epochs of the universe, one may infer that $M_{\rm max}$ has not increased significantly from $z$ = 6.42 to $z$ = 0, or from 0.85 Gyr to 13.7 Gyr after the big bang. Indeed, although Eddington-limited baryonic matter accretion is essential for seeing the quasars, this is a transient phenomenon with an estimated total lifetime of a few tens of Myr \cite{com4},
that presumably contributes very little to the final mass of the most massive black holes today.\\[.2cm]
In summary, it seems that a consistent theory of the formation of the supermassive black holes should be able to explain both observational facts, the nearly time-independent upper and lower mass limits of the supermassive black holes, as well as the early formation of the most massive black holes 
around $z$ = 6.42. 
As both the baryonic matter accretion, as well as the black hole merger scenarios, do not seem to provide us with a satisfactory description of these observational facts, we are led to ask the pertinent question, whether an alternative scenario, based on the accretion of mainly dark matter, may do better. 
Galaxies are, indeed, dominated by dark matter, and as part of this dark matter is concentrated in the galactic centers, it may very well have contributed to the formation of the supermassive black holes. Of course, in order to make definite predictions, one needs to focus on a well-defined and consistent dark matter candidate. 
Thus, in section 2, we discuss the physical and cosmological properties of our sterile neutrino dark matter candidate, while in section 3 we explore an astrophysical consequence of this dark matter scenario, namely the formation of self-gravitating supermassive degenerate sterile neutrino balls, with masses between $10^{6}M_{\odot}$ and
$3 \times 10^{9}M_{\odot}$. We then discuss how these preformed neutrino balls convert efficiently and anti-hierarchically into supermassive black holes, using stellar-mass seed black holes as catalysts.
In section 4, we describe the dynamics of the accretion of degenerate sterile neutrinos onto a black hole in a simple nonrelativistic Thomas-Fermi field theory, based on the Lane-Emden equation, while our conclusions are presented in section 5.
\subsection*{2. Physics and cosmology of sterile neutrinos}
Recently, a set of three right-handed sterile neutrinos has been consistently embedded
in a renormalizable extension of the minimal standard model of particle physics, dubbed the 
$\nu$MSM \cite{asa9}.
While two of these sterile neutrinos are unstable, having masses in the 1 GeV/$c^{2}$ to 20 GeV/$c^{2}$ range, the third one, with mass around
10 keV/$c^{2}$, is a promising quasi-stable dark matter candidate. This light sterile neutrino interacts with the standard model particles only through its tiny mixing with the active neutrinos. The bulk part of the light sterile neutrinos is produced $\sim 2.3\; \mu$s after the big bang, at temperatures
\begin{equation}
 T \sim 328\left(\frac{mc^{2}}{15\; {\rm keV}}\right)^{1/3}\; {\rm MeV}/k \; \;,
\end{equation} 
well ahead of the quark-gluon and chiral restoration phase transitions, through incoherent resonant \cite{shi10} and non-resonant \cite{dod11} scattering of the active neutrinos.
For 
a wide range of the parameters of the $\nu$MSM, the sterile neutrinos are generated out of thermal equilibrium, yielding a sterile neutrino mass fraction of the total mass-energy of this universe which is consistent with that of nonbaryonic dark matter. For instance, with an initial lepton asymmetry 
\begin{equation}
L_{\nu_{e}} = \frac{ n_{\nu_{e}} - n_{\bar{\nu}_{e}} }{n_{\gamma}} = 10^{-2} \; \; ,  
\end{equation}
a mixing angle of the
sterile neutrino $\nu_{s}$ with the active neutrino $\nu_{e}$ given by 
$\sin^{2} \vartheta = 10^{-13}$, 
and a sterile neutrino mass 
$m$ = 15 keV/$c^{2}$, 
the mass fraction of the sterile neutrinos produced in the early universe is indeed 
$\Omega_{\nu_{s}} \sim 0.24$ \cite{shi10}, 
i.e. equal to that of the nonbaryonic dark matter, derived from the WMAP data \cite{spe5}.\\[.2cm]
The same tiny mixing angle, which prevents the thermal equilibration and thus the overproduction of the sterile neutrino in the early universe, also renders the sterile neutrino quasi-stable
\cite{pal12}, 
with a lifetime much larger than the age of the universe. 
In fact, with our choice of the parameters, 
$\sin^{2} \vartheta = 10^{-13}$ and $m$ = 15 keV/$c^{2}$, 
the sterile neutrino decays with a lifetime  of 
\begin{equation}
\tau (\nu_{s} \rightarrow \nu_{e} \nu \bar{\nu}) = \frac{192 
\pi^{3}}{G_{F}^{2} m^{5} \sin^{2} \vartheta} = 1.21 \times 10^{19}\; {\rm yr}\; \;,
\end{equation}
predominantly into a $\nu_{e}$ and a neutrino-antineutrino pair, $\nu_{i}$ and $\bar{\nu_{i}}$, carrying the flavours $i = e, \mu$ or $\tau$ \cite{pal12}. These standard neutrinos are all virtually unobservable because their energy is too small. But there is also a subdominant radiative decay mode, with a branching ratio 
\begin{equation}
\frac{\tau \left( \nu_{s} \rightarrow \nu_{e} \nu \bar{\nu} \right)}{\tau \left( \nu_{s} \rightarrow \nu_{e} \gamma \right)} = \frac{2 \alpha}{8 \pi} = 0.784 \times 10^{-2}
\end{equation}
into a
potentially observable photon and a $\nu_{e}$ \cite{pal12}.
However, due to the smallness of its partial decay width of 
\begin{equation}
\left[\tau (\nu_{s} \rightarrow \nu_{e} \gamma)\right]^{-1} = 0.649 \times 10^{-21} {\rm yr}^{-1}\; \; ,
\end{equation}
these photons of energy $mc^{2}$/2, which may be the ``smoking gun'' of the sterile neutrino, are  difficult to observe as well. In fact, for the chosen model parameters, a sterile neutrino dark matter concentration of mass $M$ has a luminosity of merely 
\begin{equation}
L_{X} = \frac{Mc^{2}}{2 \tau (\nu_{s} \rightarrow \nu_{e} \gamma)} = 1.84 \times 10^{25} (M/M_{\odot})\;\; {\rm erg/s} \; \; ,
\end{equation}
in photons of $mc^{2}$/2 = 7.5 keV energy. Thus the best places to look for these photons are the diffuse extragalactic X-ray background, as well as the X-rays emitted by large galaxy clusters, low-surface-brightness and dwarf galaxies that are dominated by nonbaryonic dark matter
\cite{vie13}.
For an initial lepton asymmetry $L_{\nu_{e}} \sim 10^{-10}$, which is of the order of the baryon asymmetry \cite{spe5} 
\begin{equation}
B = \frac{n_{b}}{n_{\gamma}} \sim 6\!\times\!10^{-10} \; \;,
\end{equation}
mainly non-resonant neutrino scattering contributes to the production of sterile neutrino dark matter. These sterile neutrinos inherit a nearly thermal energy spectrum from the active neutrinos
\cite{dod11}, which allows them to play the role of warm dark matter in the large-scale structure of the universe, the clusters of galaxies and the galactic halos.
They may also erase the undesirable excessive substructure on the galactic scales. \\[.2cm]
The initial lepton asymmetry does not need to be of the same order of magnitude as the baryon asymmetry. However, for a 
larger initial lepton asymmetry, like $L_{\nu_{e}} \sim 10^{-2}$, there is, in addition to non-resonant neutrino scattering, also resonant or matter-enhanced neutrino scattering contributing to the production of dark matter \cite{shi10}. The latter yields cool sterile neutrinos that have a distorted quasi-degenerate spectrum, with an average energy of about two-thirds of that of the warm sterile neutrinos, due to the resonant Mikheyev-Smirnov-Wolfenstein (MSW) oscillations
\cite{mik14}.\\[.2cm] 
A relatively large initial lepton asymmetry of
$L_{\nu_{e}} \sim 10^{-1}$ to 10$^{-2}$ is also what seems to be required, 
to bring the observed light element abundances in line with the number of three active neutrinos at nucleosynthesis \cite{ste15}. Sterile neutrinos may as well be responsible for the pulsar kicks of up to $\sim$ 1600 km/s, which magnetars acquire in supernova explosions \cite{kus16}. 
Thus our
sterile neutrino meets all the constraints which 
an acceptable dark matter particle must fulfil \cite{shi10},\cite{vie13}, but it has several remarkable additional properties that make it a rather unique candidate for dark matter.

\subsection*{3. A symbiotic black hole formation scenario}
For our model parameters $m$ = 15 keV/$c^{2}$, $\sin^{2} \vartheta = 10^{-13}$ and $L_{\nu_{e}} = 10^{-2}$, cool (or resonant) dark matter dominates over warm (or non-resonant) dark matter by a factor of about three \cite{shi10}. The cool sterile neutrinos become nonrelativistic $\sim\;22\;$min after the big bang, well after nucleosynthesis, and they begin, together with the warm sterile neutrinos and baryonic matter, to dominate the expansion of the universe $\sim$ 79 kyr after the big bang, well ahead of recombination. Thus the primordial density fluctuations of sterile neutrino dark matter have enough time to grow nonlinear and form degenerate sterile neutrino balls
\cite{mar17}, through a process called gravitational cooling \cite{bil18}, prior to the appearance of the first quasars, 850 Myr after the big bang. This collapse process may start ahead of reionization, perhaps as early as $\sim$ 320 Myr after the big bang. Initially, the free-falling sterile neutrino dark matter, dominating baryonic matter by about a factor of six \cite{spe5}, drags the baryonic matter  along towards the center of the collapse. The baryonic gas will get heated, reionized and evaporated, but the free fall of the neutrinos will not be inhibited by the Eddington radiation limit.
Eventually, the quasi-degenerate
sterile neutrino dark matter hits, $\sim$ 640 Myr after the big bang, the degeneracy pressure, bouncing off a number of times,
while ejecting a fraction of the dark matter at every bounce. The neutrino ball finally settles in a condensate of degenerate sterile neutrino matter at the center of the collapsed object, as has been shown in calculations based on time-dependent Thomas-Fermi mean field theory \cite{bil18}.\\[.2cm] 
The smallest mass that may collapse is the mass contained within the free-streaming length at matter-radiation equality, $\sim$ 79 kyr after the big bang. For $m$ = 15 keV/$c^{2}$, this 
free-streaming mass is 
$M_{\rm warm} \sim 7 \times 10^{6} M_{\odot}$ 
in the case of warm (or non-resonant) sterile neutrinos, and
$M_{\rm cool} \sim 2 \times 10^{6} M_{\odot}$ 
in the case of the dominant cool (or resonant) sterile neutrinos \cite{shi10}.
As part of the neutrino dark matter is ejected during the collapse process, the
minimal mass of a degenerate sterile neutrino ball may be somewhat smaller than
$M_{\rm cool}$, perhaps
$M_{\rm min} \sim 10^{6} M_{\odot}$,
consistent with the lower mass limit of the observed supermassive black holes.\\[.2cm]
The maximal mass that a self-gravitating degenerate neutrino ball can support gravitationally, is the Oppenheimer-Volkoff (OV) limit \cite{opp19}
\begin{equation}
M_{\rm max} = 0.5430 \; M_{\rm Pl}^{3} \; m^{-2} \; g^{- 1/2}\; \;.
\end{equation}
For $m$ = 15 keV/$c^{2}$ and spin degeneracy factor $g = 2$, this mass
$M_{\rm max} = 2.789 \times 10^{9} M_{\odot}$ is consistent with that of the
most massive black holes observed in our universe \cite{sch1},\cite{wil2}.
As such a supermassive degenerate neutrino ball has
a radius of only 4.45 Schwarzschild radii \cite{opp19},
it is almost a black hole.
Thus the maximal mass scale of these objects may be linked to the existence of a sterile neutrino of $\sim$ 15 keV/$c^{2}$ mass, in a similar fashion as the maximal mass scale of the neutron stars is linked to the effective mass of the neutron \cite{baa20}.\\[.2cm]  
Since the gravitational potential in the interior of a neutrino ball is nearly harmonic,
these objects, in particular those near the upper mass limit, are ideal breeding
grounds for stars of mass $M \simgt 25 M_{\odot}$. As soon as such a central massive star is formed
from a collapsing molecular hydrogen cloud that was attracted to the neutrino ball, 
it may be kicked out through close binary encounters 
with intruding stars. 
However, before that happens, a star of 25 $M_{\odot}$ will evaporate large portions of its hydrogen and helium envelope and become a Wolf-Rayet star. And about 3 Myr after its formation, the star undergoes a core collapse supernova explosion of type Ic, leaving a black hole of 3 to 4 $M_{\odot}$ at the center of the neutrino ball. Some of these most massive supernova explosions, occurring in high-mass neutrino balls, may be observable as long-duration $\gamma$-ray bursts \cite{che21}. As in contrast to
pulsars, black holes do presumably not acquire ``black hole kicks'' during a supernova explosion, the velocity of the stellar-mass black hole will be small compared to the escape velocity from the center of the neutrino ball.
For a ball of 3$\times$10$^{6} M_{\odot}$ mass and 25 light-days radius \cite{mar17}, 
consisting of degenerate sterile neutrinos of 15 keV/$c^{2}$ mass and degeneracy factor $g = 2$, the escape velocity from the center is 1700 km/s, 
while for a ball of the same sterile neutrinos at the OV-limit, with 2.8$\times$10$^{9}$ $M_{\odot}$ mass and 1.4 light-days radius, the 
escape velocity from the center is the velocity of light \cite{opp19}.\\[.2cm] 
The supernova explosion of the massive star,
giving birth to a stellar-mass black hole at the center, sparks the rapid
growth of the black hole through nearly radiationless, and therefore, 
Eddington-unlimited accretion of mainly degenerate sterile neutrino dark matter from the surrounding 
neutrino ball, until the supplies dry up. 
In this symbiotic scenario, the anti-hierarchical formation of the bright quasi-stellar objects and low-luminosity active galactic nuclei may be explained by the fact that the escape velocity from the center of a 3 $\times 10^{6} M_{\odot}$ neutrino ball, is 176 times smaller than that of a 
3 $\times 10^{9} M_{\odot}$ neutrino ball. The low-mass neutrino ball may, therefore, have difficulty capturing a molecular hydrogen cloud that is able to produce a massive star. In particular, a  low-mass neutrino ball may experience a large number of unsuccessful attempts, leading to 
ordinary low-mass stars, or neutron stars after supernova explosion, prior to delivering the expected stellar-mass black hole. These unwanted stellar-mass objects will eventually be ejected from the neutrino ball through pulsar kicks or close binary encounters with intruder stars from the surrounding star cluster, thus clearing the scene for the next attempt at forming this stellar-mass black hole. The randomness of this process  may very well delay the formation of a stellar-mass black hole at the center of a low-mass neutrino ball by several Gyr,
while a neutrino ball at the top of the mass scale may easily deliver the stellar-mass black hole on its first attempt in less than 10 Myr. Of course, these sketchy ideas will have to be tested in  realistic numerical simulations. However, if this scenario is correct, some low-mass neutrino balls may still be around at some galactic centers. For instance, a 10$^{6} M_{\odot}$ neutrino ball would reveal itself through its X-ray emission of 2 $\times$ 10$^{31}$ erg/s at 7.5 keV, for our model parameters. 

\subsection*{4. Accretion of a neutrino halo onto a black hole}
A particle that is initially at rest at the surface of a $3 \times 10^{6} M_{\odot}$ neutrino ball reaches the center in the free-fall time $\tau_{F} \sim 35$ yr. This is also the time frame in which the accretion process onto the central black hole reaches a steady-state flow. In the steady-state approximation, the flow is governed by Bernoulli's eq.
\begin{equation}
\phi(r) + \frac{1}{2}(u^{2}(r) + v_{F}^{2}(r)) = \phi(r_{H}) = {\rm const} \; \; . 
\end{equation}
Here $u(r)$ is the flow velocity of the infalling degenerate sterile neutrino fluid,
$v_{F}(r)$ its Fermi velocity, $\phi(r)$ the gravitational potential and $r_{H}$ the radius of the halo. Assuming that the flow makes the gravitational potential $\phi$ extremal for all values of the radius $r$, with respect to variations that satisfy the constraint of mass conservation,
\begin{equation}
\rho u = \frac{m^{4}\;g\;v_{F}^{3}\;u}{6 \pi^{2} \hbar^{3}} = {\rm const}\; \; ,
\end{equation}
we obtain
\begin{equation}
u^{2}(r) = \frac{1}{3}v_{F}^{2} (r) = c_{S}^{2} (r) \; \;,
\end{equation}
which means that the inflow of the sterile neutrinos is trans-sonic, i.e. it flows at the local velocity of sound $c_{S} (r)$. Thus Bernoulli's eq. is now
\begin{equation}
\frac{2}{3} v_{F}^{2} (r) = \phi (r_{H}) - \phi (r) = GM_{\odot} \frac{v(x)}{bx}\; \;,
\end{equation}
which defines the quantity $v(x)$. Using Poisson's eq., one can readily verify that $v(x)$ fulfils the Lane-Emden eq.
\begin{equation}
\frac{1}{x} \frac{d^{2}v(x)}{dx^{2}} = - \left( \frac{v(x)}{x} \right)^{3/2} \; \; ,
\end{equation}
provided the length scale is
\begin{equation}
b = \frac{4}{3} \; \left( \frac{3 \pi \hbar^{3}}{4 \sqrt{2} gm^{4} G^{3/2} M_{\odot}^{1/2}} \right)^{2/3} \; \; .
\end{equation}
Thus for $m$ = 15 keV/$c^{2}$ and $g = 2$ we have $b = 2.587$ lyr. The
total mass enclosed within a radius $r = bx$ is \cite{vio22}
\begin{equation}
M(x) = M_{\odot} \left[v (x) - x v\,' (x)\right]\; \;.
\end{equation}
Various solutions $v(x)$ of the Lane-Emden eq.(13), all having the total mass $M$ = 2.714 $M_{\odot}$, are shown in Fig.1. 
\begin{figure}[h]
\begin{center}
\mbox{} \hspace{-0.843cm}
\epsfysize 7.5cm
\epsfbox{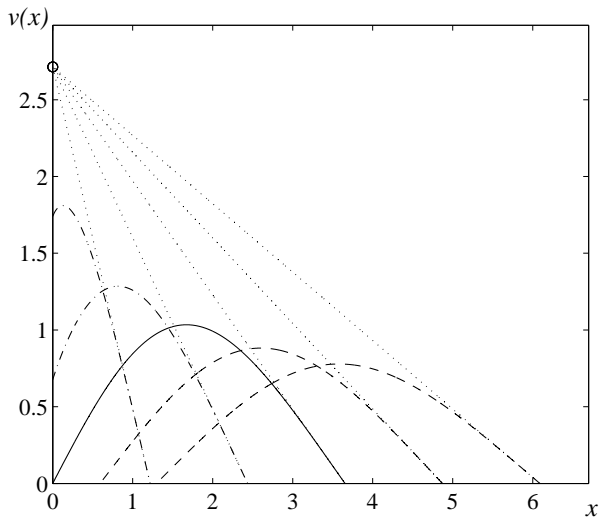}
\caption{\footnotesize Various solutions of the Lane-Emden equation, all having total mass 
$M = M_{P} + M_{H} = 2.714\;M_{\odot}$. The solid, dashed and dash-dotted lines represent the $E-$, $F-$ and $M-$ solutions, respectively.}
\end{center}
\end{figure}
There are three distinct classes of solutions.
The M-solutions exhibit a central point mass $M_{P} = M_{\odot} v$(0), surrounded by a self-gravitating degenerate sterile neutrino halo. The F-solutions describe shells of self-gravitating degenerate neutrino matter that are gravitationally unstable. The E-solution, with
$v (0) = 0$ and $v\,'(0) = 1$, stands for a pure neutrino ball with $M_{P} =M_{\odot} v$(0) = 0.
Our focus is on the M-solutions of the Lane-Emden eq.(13) because these describe, in the steady-state approximation, the various stages of the accretion history of a black hole surrounded by a degenerate sterile neutrino halo.
For instance, decreasing the halo radius $r_{H} = bx_{H}$, while keeping the total mass $M$ fixed, causes the central point mass $M_{P}$ to increase, as seen in Fig.1. The solutions of the Lane-Emden eq.(13) with $M_{P} = M_{\odot}v$(0) $>$ 0,
for arbitrary mass $\simM$ can be obtained noting that, if $v(x)$ is a solution, 
$\simv(x)=A^{3} v(Ax)$ with $A > 0$ is a solution as well.
Thus all the masses and radii scale as $\simM=A^{3} M$ and
$\simr=r/A$ \cite{vio22}. The mass accretion rate into a sphere, containing the mass $M_{C}$ within a radius $r_{C}$ from the center, is thus given by 
\begin{equation}
\frac{dM_{C}}{dt} = 4 \pi r_{C}^{2} \rho (r_{C}) u (r_{C})
\end{equation}
 or
\begin{equation}
\frac{d M_{C}}{dt} \; = \; \frac{ \sqrt{3} g m^{4} G^{2} M_{\odot}^{2}}{2 \pi \hbar^{3}} \; 
\left[ v(x_{C}) \right]^{2} \; = \;
\frac{f(\mu)}{\tau M_{\odot}} \; M_{C}^{2} \; \; ,
\end{equation}
where we have introduced the universal time-scale
\begin{equation}
\tau = \frac{2 \pi \hbar^{3}}{\sqrt{3} g m^{4} G^{2} M_{\odot}} =
1.488\!\times\!10^{7} \; \frac{2}{g} \; 
\left( \frac{15\;{\rm keV}}{mc^{2}} \right)^{4} \; {\rm yr} \; \; .
\end{equation} 
The shut-off parameter
\begin{equation}
f(\mu) = \frac{v \left( x_{C} \right)^{2}}{ \left[ v (x_{C}) - x_{C} v\,' (x_{C}) \right]^{2}}
\end{equation}
is a function of the mass ratio $\mu = M_{C}/M$.
We now choose $r_{C} = bx_{C}$ to be the radius at which the 
escape velocity reaches the velocity of light, i.e. $x_{C}$ is given by
\begin{equation}
\frac{1}{x_{\odot}} = \frac{v (x_{C})}{x_{C}} - v\,'(x_{H})\; \; ,
\end{equation}
where $r_{\odot} = bx_{\odot}$ is the Schwarzschild radius of the sun with
\begin{equation}
x_{\odot} = \frac{2 GM_{\odot}}{bc^{2}} =
{\rm 1.207} \times 10^{-13}
\; \left( \frac{g}{2} \right)^{2/3} \;
\left( \frac{mc^{2}}{15\;{\rm keV}} \right)^{8/3} \; \; .
\end{equation}
\begin{figure}[h]
\begin{center}
\mbox{} \hspace{1cm}
\epsfysize 7.5cm
\epsfbox{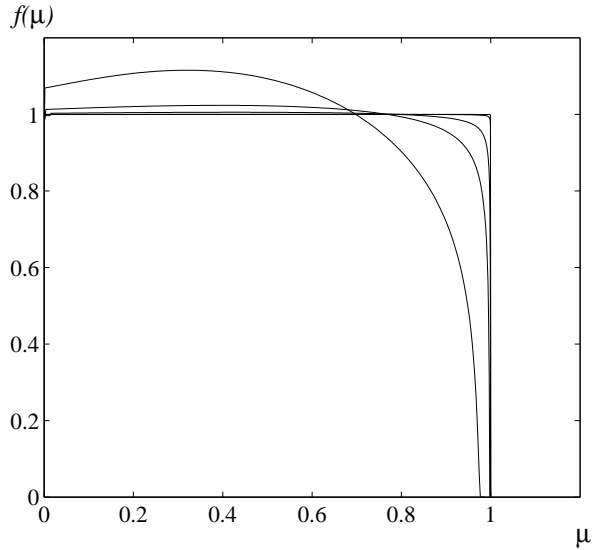}
\caption{\footnotesize Various shut-off parameters $f(\mu)$ as a function of the mass ratio $\mu = M_{C}/M$. The neutrino ball masses vary from $M = 10^{6} M_{\odot}$ for the box function, through $M = 10^{7}, 10^{8}$ to 10$^{9}\;M_{\odot}$ for the curve with the largest peak value.}
\end{center}
\end{figure}
As for $M \ll M_{\rm max}$ and $\mu < 1$, we may approximate $v (x_{C}) \sim v$(0) and
$f( \mu) \sim 1$, eq.(17) agrees well with standard Bondi accretion theory \cite{bon23}. Integrating eq.(17) using this approximation, the growth of the black hole is given by
\begin{equation}
M_{C} (t) \sim \frac{M_{C}(0)}{1 - t/ \tau_{A}}\; \; ,
\end{equation} 
yielding an
accretion time-scale of
$\tau_{A} = \tau M_{\odot}/M_{C}$(0).
During the accretion process, both $M_{C}$ and $x_{C}$ grow, while $x_{H}$
shrinks as a function of time, eventually causing $x_{C}$ and $x_{H}$ to converge and $v(x_{C})$ to vanish.
The shut-off parameter $f(\mu)$, shown in Fig.2 as a function of the mass ratio $\mu = M_{C}/M$, is for a neutrino ball of mass $M = 10^{6}\;M_{\odot}$ a simple Heavyside function. As $M$ increases towards $M_{\rm max}$, this curve starts deviating from the simple box form, thus signalling
the breakdown of our 
nonrelativistic theory. 
Since, for $M \ll M_{\rm max}$, the black hole growth is approximately given by the Bondi formula, $\dot{M}_{C} \propto M_{C}^{2}$, 
we expect the mass growth curves to match the Bondi solution closely,
with the attractive feature that the solution of eq.(17) eventually brings the growth to a halt. 
The mass ratio $\mu$ is shown in Fig.3 as a function of time in units of the accretion time-scale 
$\tau_{A}$.
For low-mass central black holes, these curves are indistinguishable from the Bondi solution, 
represented by the dotted curve. Here we start with $\mu(0) = 0.1$ at $t = 0$ to illustrate the differences for halo masses between $10^{6}$ and $10^{9} M_{\odot}$.
The universal time-scale is $\tau$ = 14.88 Myr, and the accretion time-scales for a $3\!\times\!10^{6} M_{\odot}$
neutrino ball onto 3 and 4 $M_{\odot}$ seed black holes are, therefore, $\tau_{A} = 4.96$ Myr and 
$\tau_{A} = 3.72$ Myr, respectively.
\begin{figure}[h]
\begin{center}
\mbox{} \hspace{-0.94cm}
\epsfysize 7cm
\epsfbox{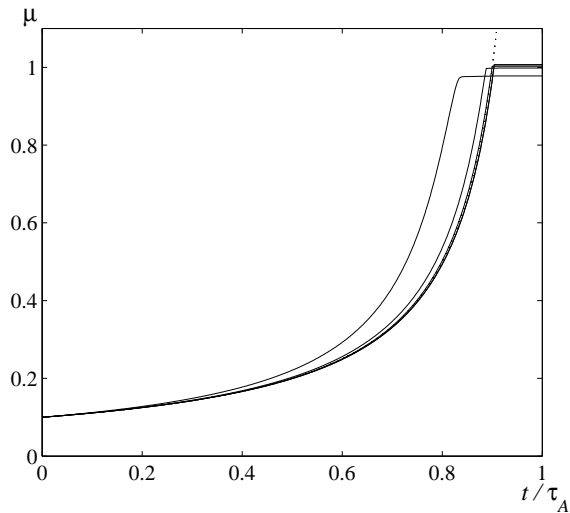}
\end{center}
\caption{\footnotesize The growth of a central black hole as a function of time, in units of the accretion time $\tau_{A}$, for halos of between $10^{6}$ and $10^{9} M_{\odot}$, as in Fig.2. For $\mu < 0.1$, the growth curves match the Bondi growth curve too closely to distinguish.} 
\end{figure}
\subsection*{5. Conclusions}
In summary, the neutrino balls are almost entirely swallowed by the seed black holes in an accretion 
time scale  
$\tau_{A} <$ 5 Myr, thus converting these rapidly into supermassive 
black holes with negligible residual sterile neutrino halos. 
The most massive black holes may, therefore, form between 650 Myr and 840 Myr after the big bang. Although the nonrelativistic Thomas-Fermi mean field theory breaks down for $M \sim M_{\rm max}$ and $r_{H} \sim r_{C}$, we expect these basic results to persist in a relativistic theory of the accretion process.\\[.2cm]
There are three main features which distinguish the accretion of degenerate sterile neutrino dark matter
from that of baryonic matter, playing a decisive role in the rapid growth of a stellar-mass black hole to the supermassive scale.
Firstly, in contrast to the clumping of baryonic matter, neither the neutrino ball formation nor the neutrino halo accretion onto a black hole is inhibited by the Eddington radiation limit.
Secondly, the matter densities of the 
degenerate sterile neutrino balls are much larger than those of any known form of baryonic matter having the same total mass, leading to much faster growth of the black holes through neutrino dark matter.
Thirdly, the preformed degenerate sterile neutrino balls have, for $m \sim$ 15 keV/$c^{2}$ and $g$ = 2, masses in the same range as the observed supermassive black holes, which sets a natural limit to the growth of the black holes.\\[.2cm] 
We, therefore,
conclude that supermassive neutrino balls, with stellar-mass black holes at their center, indeed offer an intriguing symbiotic scenario, in which baryonic matter conspires with degenerate sterile neutrino dark matter, to form these galactic supermassive black holes, with masses between 10$^{6}\;M_{\odot}$ and $3 \times 10^{9}\;M_{\odot}$, rapidly and efficiently.

\newpage


\noindent
{\large \bf Acknowledgements}\\[.2cm]
\noindent
This research is supported by the Foundation for Fundamental Research and the National Research Foundation of South Africa. 



\scriptsize

\begin{thebibliography}{99}
\bibitem{sch1} M. Schmidt, 1963
               {\it Nature} {\bf 197} 1040\\
               E.E. Salpeter,  
               1964 {\it Astrophys. J.} {\bf 140} 796\\
                Y.B. Zel'dovich, 1964  
               {\it Dokl. Akad. Nauk SSSR} {\bf 155} 67\\
               D. Lynden-Bell, 1969  
               {\it Nature} {\bf 223} 690\\
               J. Kormendy and D. Richstone, 1995 
               {\it Ann. Rev. Astron. Astrophys.} {\bf 33} 581\\
               J. Kormendy and L.C. Ho,  
               astro-ph/0003268\\
                L.C. Ho and J. Kormendy,
               astro-ph/0003267.
\bibitem{wil2} C.J. Willott, R.J. McLure and M.J. Jarvis, 2003 
                 {\it Astrophys. J.} {\bf 587} L15\\
               F. Walter {\it et al}., 2003 
               {\it Nature} {\bf 424} 406.
\bibitem{ell3} R. Ellis, 1998 
                {\it Nature} {\bf 395} A3-8.
\bibitem{com4} F. Combes, 
                astro-ph/0505463.
\bibitem{spe5} D.N. Spergel {\it et al.}, 
                astro-ph/0603449.
\bibitem{hil6} J.G. Hills, 1975
                {\it Nature} {\bf 254} 295\\
                M.J. Rees, 1984 
                {\it Ann. Rev. Astro. Astrophys.} {\bf 22} 471.
\bibitem{mil7} M.C. Miller and E.J.M. Colbert, 2004
                {\it Int. J. Mod. Phys.} {\bf D13} 1.
\bibitem{mac8} F. Macchetto {\it et al}., 1997
                  {\it Astrophys. J.} {\bf 489} 579.
\bibitem{asa9} T. Asaka, S. Blanchet and M. Shaposhnikov, 2005
                {\it Phys. Lett.} {\bf B631} 151\\
                T. Asaka and M. Shaposhnikov, 2005
                {\it Phys. Lett.} {\bf B620} 17\\
              M. Shaposhnikov and I. Tkachev, 
                hep-ph/0604236.
\bibitem{shi10} X. Shi and G.M. Fuller, 1999
                {\it Phys. Rev. Lett.} {\bf 82} 2832\\
                K. Abazajian, G.M. Fuller and M. Patel, 2001 
                {\it Phys. Rev.} {\bf D64} 023501\\
                K.N. Abazajian and G.M. Fuller, 2002 
                {\it Phys. Rev. } {\bf D66} 023526\\
                K. Abazajian and S.M. Koushiappas,  
                astro-ph/0605271.
\bibitem{dod11} S. Dodelson and L.M. Widrow, 1994  
                {\it Phys. Rev. Lett.} {\bf 72} 17\\
                R. Barbieri and A.D. Dolgov, 1990 
                {\it Phys. Lett.} {\bf B237}  440\\
                 K. Kainulainen, 1990 
                {\it Phys. Lett.} {\bf B244}  191.
\bibitem{pal12} P.B. Pal and L. Wolfenstein, 1982  
                {\it Phys. Rev.} {\bf D25}  766.
\bibitem{vie13} M. Viel, J. Lesgourgues, M.G. Haehnelt, S. Matarrese and A. Riotto, 
                astro-ph/0605706\\
%
                A. Boyarski, A. Neronov, O. Ruchayskiy, M. Shaposhnikov and I. Tkachev, 
                astro-ph/0603660.
\bibitem{mik14} S.P. Mikheyev and A. Yu. Smirnov, 1985
                {\it Sov. J. Nucl. Phys.} {\bf 42} 913\\
L. Wolfenstein, 1978 
                {\it Phys. Rev.} {\bf D17} 2369.
\bibitem{ste15} G. Steigman, 2006  
                {\it Int. J. Mod. Phys.} {\bf E15}  1.
\bibitem{kus16} A. Kusenko and G. Segr\'{e}, 1997  
                {\it Phys. Lett.} {\bf B396} 197\\
                 A. Kusenko and G. Segr\'{e}, 1999  
                {\it Phys. Rev.} {\bf D59}  061302.
\bibitem{mar17} M.A. Markov, 1964
                {\it Phys. Lett.} {\bf 10}  122\\
               R. Ruffini, 1980 
                {\it Lett. Nuovo Cim.} {\bf 29}  161\\
                 R.D. Viollier, D. Trautmann and G.B. Tupper, 1993
                {\it Phys. Lett.} {\bf B306} 79\\
                F. Munyaneza, D. Tsiklauri and R.D. Viollier, 1998
                {\it Astrophys. J.} {\bf 509} L105\\
               F. Munyaneza, D. Tsiklauri and R.D. Viollier,  1999
                {\it Astrophys. J.} {\bf 526}  744.
\bibitem{bil18} N. Bili\'{c}, R.J. Lindebaum, G.B. Tupper and R.D. Viollier, 2001 
                {\it Phys. Lett.} {\bf B515} 105.
\bibitem{opp19} J.R. Oppenheimer and G.M. Volkoff, 1939  
                {\it Phys. Rev.} {\bf 55}  374\\
                N. Bili\'{c}, F. Munyaneza and R.D. Viollier, 1999  
                {\it Phys. Rev.} {\bf D59} 024003.
\bibitem{baa20} W. Baade and F. Zwicky, 1934
                {\it Phys. Rev.} {\bf 45}  138\\
              A. Hewish, S.J. Bell, J.D.H. Pilkington, P.F. Scott, and R.A. Collins, 
             1968
                {\it Nature} {\bf 217} 709.
\bibitem{che21} R.A. Chevalier and Z. Li, 1999 
                {\it Astrophys. J.} {\bf 520}  L29\\
                P.A. Mazzali {\it et al}.,
                astro-ph/0603567.
\bibitem{vio22} R.D. Viollier, F.R. Leimgruber and D. Trautmann,  1992
                {\it Phys. Lett.} {\bf B297}  132.
\bibitem{bon23} H. Bondi, 1952 
                {\it Mon. Not. Roy. Astron. Soc.} {\bf 112} 195.
\end{thebibliography}
\end{document}